\documentclass[letter,10pt]{IEEEtran}

\usepackage[utf8]{inputenc}
\usepackage{amsmath,amsthm,amssymb,epsfig,mathrsfs,mathtools,cite,afterpage,color}
\usepackage{tikz,pgfplots,pgfplotstable}
\usetikzlibrary{shapes,arrows,trees}
\usetikzlibrary{decorations.pathreplacing}
\usepackage{verbatim}
\usepackage[linktoc=none]{hyperref}
\newtheorem{theorem}{Theorem}[section]
\newtheorem{lemma}{Lemma}[section]
\newtheorem{definition}{Definition}[section]
\newtheorem{remarks}{Remarks}[section]
\newtheorem{corollary}{Corollary}[section]

\def\DEFARXIV{1}
\def\DEFMAIN{1}
\def\DEFAPPENDIX{0}

    \tikzstyle{block} = [draw, fill=blue!20, rectangle, minimum height=1.5em,minimum width=2em,node distance=2.2cm]
    \tikzstyle{sum} = [draw, fill=blue!20, circle, node distance=2.2cm]
    \tikzstyle{input} = [coordinate]
    \tikzstyle{output} = [coordinate]
    \tikzstyle{pinstyle} = [pin edge={to-,thin,black}]
    \tikzstyle{connector} = [->,thick,above]

\ifnum\DEFARXIV=0
\author{
\IEEEauthorblockN{Author 1, Author 2, Author 3}
}
\else
\author{
\IEEEauthorblockN{Sundaram R M, Devendra Jalihal, Venkatesh Ramaiyan}
\IEEEauthorblockA{Department of Electrical Engineering,\\
Indian Institute of Technology Madras,\\
Chennai 600036, India.\\
Email: sundaram.rm@gmail.com, \{dj, rvenkat\}@ee.iitm.ac.in}
}
\fi

\begin{document}
\ifnum\DEFMAIN=1
\ifnum\DEFARXIV=0
	\title{ Optimal Sequential Frame Synchronization in a Fading Channel}
\else
	\title{ Asynchronous Communication over a Fading Channel and Additive Noise }
\fi
\maketitle
\begin{abstract}
In \cite{Chandar2008}, Chandar et al studied a problem of sequential frame synchronization for a frame transmitted
randomly and uniformly among $A$ slots. For a discrete memory-less channel (DMC), they showed that the frame length
$N$ must scale as $e^{\alpha(Q)N}>A$ for the frame detection error to go to zero asymptotically with $A$. $\alpha(Q)$
is the synchronization threshold and $Q$ is channel transition probability.
We study the sequential frame synchronisation problem for a fading channel and additive noise 
and seek to characterise the effect of fading.
For a discrete ON-OFF fading channel (with ON probability $p$) and additive noise (with
channel transition probabilities $Q_n$),
we characterise the synchronisation threshold of the composite channel $\alpha(Q)$ and show that $\alpha(Q)\leq p\,\alpha(Q_n)$.
We then characterize the synchronization threshold for Rayleigh fading and AWGN channel as a function of channel parameters.
The asynchronous framework permits a trade-off between sync frame length, $N$, and channel, $Q$,
to support asynchronism. This allows us to characterize the synchronization threshold with sync frame
energy instead of sync frame length.
\end{abstract}

\section{Introduction}
Frame synchronization generally
concerns the problem of identifying the sync word, which points the start of a frame, imbedded in a continuous stream of framed data (see e.g., \cite{Massey1972}).
The problem of detecting and decoding data transmitted sporadically is studied as asynchronous communication.
For example, 
the objective of the asynchronous
communication
system could be to detect and decode a single frame transmitted at some random and unknown time
and there may be no transmission before or after the frame (see e.g., \cite{Tchamkerten2009}).

The problem of asynchronous communication has been studied earlier in works
such as \cite{Massey1972} and \cite{Mehlan1993}, but the interest
has increased in recent times with emerging applications in wireless sensor networks and some control channels.
In wireless sensor and actor networks (see e.g., \cite{Akyildiz2004} and \cite{Akyildiz2002}),
the participating nodes would report a measurement or an event to the fusion centre at random epochs.
The nodes may need to transmit few bytes of data to the fusion centre over a
relatively large time frame, e.g., a single packet possibly in an hour or even in a day.
Also, in frameworks such as the Internet of Things \cite{IoT_Atzori2010}, the nodes
may report measurements sporadically leading to an asynchronous communication framework,
however, the constraint on power may be less stringent than in wireless sensor networks.
Characterisation of the communication overheads needed in such setup is crucial
for optimal network design and operation.

\subsection{Related Literature}
In \cite{Chandar2008}, Chandar et al studied a problem of sequential frame synchronization
for a frame transmitted randomly and uniformly in an interval of known size. For a discrete memory-less channel,
they defined a synchronisation threshold that characterises the sync frame length needed for
error-free frame detection.
In our work, we study the sequential frame
synchronisation problem for a fading channel and seek to characterise
the effect of fading. 
In \cite{Tchamkerten2009}, a
basic
framework for communication in an asynchronous set up is proposed and
achievable trade-off between reliable communication and asynchronism is discussed.
The asynchronous communication set-up is studied in the finite block-length regime in \cite{Polyanskiy2013}.
We restrict our attention to the frame synchronisation problem
and study a generalisation that permits us to characterize the scaling
needed of the sync frame energy for asymptotic error-free frame synchronisation
(instead of the sync frame length $N$ considered in \cite{Chandar2008} and \cite{Tchamkerten2009}). 


\section{Problem Set-up}
\label{sec:setup}
\begin{figure}
\begin{center}
\begin{tikzpicture}[scale=3]
	\clip(-0.1,0.35) rectangle(2.1,.75);
	\foreach \i in {0,...,20} {
	  \draw[very thin] (\i/10,.5-0.02) -- (\i/10,.5+.02);
	  }
	\draw (.0,.5) node [below]{$0$} -- (2,.5) node [below]{$A$};
	\foreach \i in {1,...,4} {
	 \path (\i/10,0.5) node [font=\scriptsize,above] {$\cdotp$};
	}
	\draw [decorate,decoration={brace}]
(1/10,0.5+0.1) -- (4/10,0.5+0.1) node [midway,above]{ \footnotesize $x(0)$'s};

	\path (5/10,0.5) node [font=\scriptsize,above] {$s_1$};
	\path (5/10,0.5) node [font=\scriptsize, below] {$v$};
	\path (6/10,0.5) node [font=\scriptsize,above] {$s_2$};
	\path (7/10,0.5) node [font=\scriptsize,above] {$\cdotp$};
	\path (8/10,0.5) node [font=\scriptsize,above] {$\cdotp$};
	\path (9/10,0.5) node [font=\scriptsize,above] {$s_N$};
	\draw [thick, decorate,decoration={brace}]
(5/10,0.5+.1) -- (9/10,0.5+0.1) node [midway,above]{ \footnotesize
$\mathbf{s}^N$};

	\foreach \i in {10,...,20} {
	 \path (\i/10,0.5) node [font=\scriptsize,above] {$\cdotp$};
	}
	\draw [decorate,decoration={brace}]
(10/10,0.5+.1) -- (20/10,0.5+0.1) node [midway,above]{ \footnotesize $x(0)$'s};
\end{tikzpicture}
\end{center}
\caption{A discrete-time asynchronous communication model. A sync packet $\mathbf{s}^N = (s_1,\cdots,s_N)$
is transmitted at some random time $v \sim U\{1,A\}$.
The channel input in slots other than $\{v, \cdots, v+N-1\}$
is assumed to be $x(0)$.
}
\label{fig_async}
\end{figure}
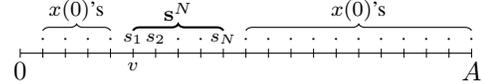

The problem set-up is illustrated in Figure~\ref{fig_async}.
We consider discrete-time communication between a transmitter
and a receiver over a discrete memory-less channel.
The discrete memory-less channel is characterized by finite input and output alphabet sets ${\mathcal X}$ and
${\mathcal Y}$ respectively, and transition probabilities $Q(y|x)$ defined for all
$x \in {\mathcal X}$ and $y \in {\mathcal Y}$.

A sync packet $\mathbf{s}^N = (s_1, \cdots, s_N)$ of length $N$ symbols
($s_i \in {\mathcal X}$ for all $i = 1,\cdots,N$)
is transmitted at some random time, $v$, distributed Uniformly in $\{1,2, \cdots, A\}$, where $A$ is assumed known.
We suppose that arrival of data from
higher layers or occurrence of an event at time $v$ triggers the transmission
of the sync packet.
The transmission occupies slots $\{v, v+1, \cdots, v + N -1\}$ as illustrated in Figure~\ref{fig_async},
i.e., $x_n = s_{n-v+1}$ for $n \in \{v, \cdots, v + N - 1\}$,
and, we assume that the channel input in slots other than $\{v, v+1, \cdots, v + N -1\}$ is
$x(0)$ ($x(0) \in {\mathcal X}$ and could represent zero input).
The distribution of the channel output, $\{y_n\}$,
conditioned on the random time $v$ and the
sync sequence ${\mathbf s}^N$, is
$Q(\cdot|s_{n-v+1})$ for $n \in \{v, v+1, \cdots, v+N-1\}$ and $Q(\cdot|x(0))$ otherwise.

The receiver seeks to identify the location of the sync packet $v$ from the channel output $\{y_n\}$.
Let $\hat{v}$ be
an estimate of $v$. Then, the error event
is represented as $\{\hat{v} \neq v\}$ and the associated probability of error in frame synchronization
would be ${\mathsf P}(\{\hat{v} \neq v\})$.
We are interested in characterizing the sync sequence ${\mathbf s}^N$ needed for asymptotic error-free
frame synchronization.
In this paper, we assume that the receiver employs a sequential decoder to detect the sync packet.
In particular, we assume that the decision $\hat{v} = t$ depends only on the output sequence up to time $t+N-1$,
i.e., $\{y_1, \cdots, y_t, \cdots , y_{t+N-1}\}$.

In \cite{Chandar2008}, Chandar et al define a synchronization threshold that characterizes
the sync frame length needed for optimal sequential
frame synchronisation for a discrete memoryless channel.
\begin{definition}[from \cite{Chandar2008}]
Let $A = e^{\alpha N}$. An asynchronism exponent $\alpha$ is said to be achievable
if there exists a sequence of pairs, sync pattern and sequential decoder $( \mathbf{s}^N,\hat{v} )$, for
all $N \geq 1$, such that
\[ {\mathsf P}(\{ \hat{v} \neq v \}) \rightarrow 0 \mbox{\ \ as \ } {N \rightarrow \infty} \]
The synchronization threshold, denoted as $\alpha(Q)$, is defined as the
supremum of the set of achievable asynchronism exponents.
\end{definition}
In \cite{Chandar2008},
the synchronization threshold for the discrete memory-less channel was shown to be
\begin{equation}
\alpha(Q) = \max_{x \in {\mathcal X}} D( Q(\cdot|x) \| Q(\cdot|x(0)) ) 
\end{equation}
where $D( Q(\cdot|x) \| Q(\cdot|x(0)) )$ is the Kullback-Leibler distance between $Q(\cdot|x)$
and $Q(\cdot|x(0))$.
The authors also
provide a construction of sync sequence $\mathbf{s}^N$
entirely with two symbols, $x(0)$ and $x(1)$, where
\[ x(1) := \arg\max_{x \in {\mathcal X}}D( Q(\cdot|x) \| Q(\cdot|x(0)) ) \]
and show asymptotic error-free frame synchronization with a sequential
joint typicality decoder\footnote{See the proof of Theorem~\ref{thm:generalization} in Section~\ref{sec:generalization}
for details on the joint typicality decoder.}.

In this work, we study the sequential frame synchronisation problem for a fading channel and seek to characterise
the effect of fading. In Section~\ref{sec:fading},
we characterise the synchronisation threshold for a composite, general fading and additive noise channel with finite
alphabets. The synchronisation threshold for the Rayleigh fading and AWGN channel is studied in Section~\ref{sec:rayleigh_fading}.
In Section~\ref{sec:generalization}, we generalise the frame synchronisation framework and study a
tradeoff between sync frame length $N$ and the channel.


\section{Composite Fading and Additive Noise Channel}
\label{sec:fading}
\begin{figure}
\begin{center}
\def\layersep{2.5cm}
\begin{tikzpicture}[draw=black!50, node distance=\layersep,font=\scriptsize]
    \tikzstyle{sym}=[circle,fill=black!20,minimum size=5pt,inner sep=0pt]
    \tikzstyle{xsym}=[sym, color=black!60,text=black];
    \tikzstyle{ysym}=[sym, color=black!40,text=black];
    \tikzstyle{dysym}=[sym, color=black!40,opacity=0.5,text=black];
    \tikzstyle{hsym}=[sym, color=black!30,text=black];
    \tikzstyle{dhsym}=[sym, color=black!30, opacity=0.5,text=black];
    \tikzstyle{annot} = [text width=4em, text centered]

    \foreach \name / \y in {1,2}
    {
        \node[xsym] (I-\name) at (0,-\y-1) {};
        \node[left of=I-\name,node distance=10] {$x_\y$};
     }

    \foreach \name / \y in {1,5}
    {
        \path[yshift=0.5cm]
            node[hsym] (H-\name) at (\layersep,-\y cm) {};
            \node[above of=H-\name,node distance=10] {$h_\y$};
    }

    \foreach \name / \y in {2,3,4}
    {
        \path[yshift=0.5cm]
            node[dhsym] (H-\name) at (\layersep,-\y cm) {};
        \node[above of=H-\name,node distance=10] {$h_\y$};
    }
            
    \foreach \name / \y in {1,3}
    {
	 \path[yshift=0.5cm]
	  node[ysym] (O-\name) at (2*\layersep,-\y cm -1 cm) {};
	 \node[right of=O-\name,node distance=10] {$y_\y$};
	}
    \foreach \name / \y in {2}
    {
	\path[yshift=0.5cm]
	  node[dysym] (O-\name) at (2*\layersep,-\y cm -1 cm) {};
	\node[right of=O-\name,node distance=10] {$y_\y$};
	}

    \foreach \source in {1,...,2}
        \foreach \dest in {1,5}
            \path [line width=1pt] (I-\source) edge node (IH-\source\dest) {} (H-\dest);
    \foreach \source in {1,...,2}
        \foreach \dest in {2,3,4}
            \path (I-\source) edge node (IH-\source\dest) {} (H-\dest);

    \foreach \source in {1,5}
        \foreach \dest in {1,3}
            \path [line width=1pt] (H-\source) edge node (HO-\source\dest) {} (O-\dest);
    \foreach \source in {1,5}
        \foreach \dest in {2}
            \path (H-\source) edge node (HO-\source\dest) {} (O-\dest);
    \foreach \source in {2,...,4}
        \foreach \dest in {1,...,3}
            \path (H-\source) edge node (HO-\source\dest) {} (O-\dest);

    \node[annot,above of=H-1, node distance=1cm] (hl) {$\mathcal{H}$};
    \node[annot,left of=hl] {$\mathcal{X}$};
    \node[annot,right of=hl] {$\mathcal{Y}$};
    \node[above of=HO-11,node distance=0.4 cm] {$Q_n(y_1|h_1)$};
    \node[below of=HO-53,node distance=0.4 cm] {$Q_n(y_5|h_3)$};
    \node[above of=IH-11,node distance=0.5 cm] {$H(h_1|x_1)$};
    \node[below of=IH-25,node distance=0.5 cm] {$H(h_5|x_2)$};
\end{tikzpicture}
\caption{A composite fading and additive noise channel containing finite alphabets. The transition probabilities for the fading channel and the additive noise channel are provided by $H(\cdot|\cdot)$ and $Q_n(\cdot|\cdot)$ respectively.}
\label{fig:composite_channel}
\end{center}
\end{figure}
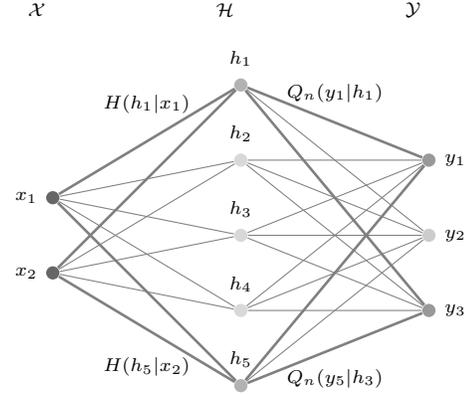
In this section, we consider a discrete, memory-less, composite fading and additive noise channel
$n(h(\cdot)): {\mathcal X} \rightarrow {\mathcal Y}$
with finite input and output alphabet sets ${\mathcal X}$ and ${\mathcal Y}$ respectively (see Figure~\ref{fig:composite_channel}).
The random fading channel $h(\cdot)$ is modelled as $h: {\mathcal X} \rightarrow {\mathcal H}$,
where ${\mathcal H}$ is assumed to be a finite alphabet set,
with transition probabilities $H(h|x)$ defined for all $x \in {\mathcal X}$ and $h \in {\mathcal H}$.
And, the additive noise channel $n(\cdot)$ is modelled as $n:{\mathcal H} \rightarrow {\mathcal Y}$
with transition probabilities $Q_n(y|h)$ defined for all $h \in {\mathcal H}$ and $y \in {\mathcal Y}$.
Further, we will assume that the fading process $h(\cdot)$ is independent of the additive noise process $n(\cdot)$.
Then, the transition probabilities of the composite channel,  $Q(\cdot|\cdot)$, is defined for
all $x \in {\mathcal X}$ and $y \in {\mathcal Y}$ as
\[ Q(y|x) = \sum_h H(h|x) Q_n(y|h) \]
From Theorem 1 of \cite{Chandar2008}, the synchronisation threshold
for the composite fading and additive noise channel is given by \begin{equation}
\alpha(Q) = \max_{x \in {\mathcal X}} D( Q(\cdot|x) \| Q(\cdot|x(0)) ) \label{eqn:alpha}
\end{equation}

We will now characterise the effect of fading by comparing $\alpha(Q)$ with $\alpha(Q_n)$ (the synchronisation threshold
for the additive noise channel). 
In the remainder of this section, we will restrict to an ON-OFF fading channel, where $h(x) \in \{ x, x(0) \}$ for
all $x \in {\mathcal X}$.
Further, we will assume that the transition probabilities for
the ON-OFF fading channel is parameterized by the ON probability $p$,
i.e.,
\begin{equation}
H(h|x) = p \ I_{\{h = x\}} + (1-p) \ I_{\{h = x(0)\}}
\label{eqn:transition_probabilities}
\end{equation}
Then, the transition probabilities for the composite channel is given by,
\[ Q(y|x) = p \ Q_n(y|x) + (1-p) \ Q_n(y|x(0)) \]
The synchronisation threshold $\alpha(Q)$ for the composite channel is now characterised
in the following lemma.
\begin{lemma}
\label{lem:fading}
$\alpha(Q) \leq p \ \alpha(Q_n)$ \qed
\end{lemma}
\ifnum\DEFARXIV=0
The proof of all lemmas and theorems is available in the appendix of the extended version of this 
paper\footnote{Provided as a supporting document}.
\fi
\ifnum\DEFARXIV=1
	\begin{proof}
	Define $x(1) := \underset{x \in {\mathcal X}}{\arg\,\max\,} D( Q(\cdot|x) \| Q(\cdot|x(0)) )$.
	Then,
	\begin{IEEEeqnarray}{rCl}
	\alpha(Q) &=& \max_{x \in {\mathcal X}} D( Q(\cdot|x) \| Q(\cdot|x(0)) ) \nonumber \\
	&=& D( Q(\cdot|x(1)) \| Q(\cdot|x(0)) ) \nonumber \\
	&=& D( p \ Q_n(\cdot|x(1)) + (1-p) \ Q_n(\cdot|x(0)) 
			\| Q_n(\cdot|x(0)) ) \nonumber \\
	&\leq& p \ D(Q_n(\cdot|x(1)) \| Q_n(\cdot|x(0)) \label{eqn:fading}
	\end{IEEEeqnarray}
	The last equation follows from Jensen's inequality. Also, we know that
	for an independent fading and additive noise model,
	\begin{IEEEeqnarray*}{rCl}
	\IEEEeqnarraymulticol{3}{l}{\arg\max_{x \in {\mathcal X}} D( Q(\cdot|x) \| Q(\cdot|x(0)) )} \\
	&=& x(1) \\
	&=& \arg\max_{x \in {\mathcal X}} D( Q_n(\cdot|x) \| Q_n(\cdot|x(0)) )
	\end{IEEEeqnarray*}
	Thus, we have, $D(Q_n(\cdot|x(1)) \| Q_n(\cdot|x(0)) = \alpha(Q_n)$ and, substituting in Equation~(\ref{eqn:fading}), we get,
	\[ \alpha(Q) \leq p \ \alpha(Q_n) \]
	\end{proof}
\fi 

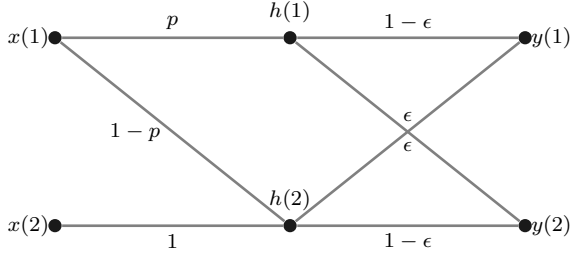
\begin{figure}
\begin{center}
\def\layersep{2.5cm}
\begin{tikzpicture}[draw=black!50, node distance=\layersep,font=\footnotesize,scale=1.25]
    \tikzstyle{sym}=[circle,fill=black!20,minimum size=5pt,inner sep=0pt]
    \tikzstyle{xsym}=[sym, color=black!90,text=black];
    \tikzstyle{ysym}=[sym, color=black!90,text=black];
    \tikzstyle{hsym}=[sym, color=black!90,text=black];

    \foreach \name / \y in {1,2}
    {
        \path[yshift=.5cm]
         	node[xsym] (I-\name) at (0,-2*\y) {};
        \node[left of=I-\name,node distance=10] {$x(\y)$};
    }

    \foreach \name / \y in {1,2}
    {
        \path[yshift=0.5cm]
            node[hsym] (H-\name) at (\layersep,-2*\y ) {};
		\node[above of=H-\name,node distance=10] {$h(\y)$};
    }
           
    \foreach \name / \y in {1,2}
    {
	 \path[yshift=0.5cm]
	   node[ysym] (O-\name) at (2*\layersep,-2*\y) {};
	 \node[right of=O-\name,node distance=10] {$y(\y)$};
	}

    \foreach \source in {1}
        \foreach \dest in {1,2}
            \path [line width=1pt] (I-\source) edge node (IH-\source\dest) {} (H-\dest);
    \path [line width=1pt] (I-2) edge node (IH-22) {} (H-2);

    \foreach \source in {1,2}
        \foreach \dest in {1,2}
            \path [line width=1pt] (H-\source) edge node (HO-\source\dest) {} (O-\dest);

    \node[above of=IH-11,node distance=0.2 cm] {$p$};
    \node[left of=IH-12,node distance=0.5 cm] {$1-p$};
    \node[below of=IH-22,node distance=0.2 cm] {$1$};
    \node[above of=HO-11,node distance=0.2 cm] {$1-\epsilon$};
    \node[above of=HO-12,node distance=0.2 cm] {$\epsilon$};
    \node[below of=HO-21,node distance=0.2 cm] {$\epsilon$};
    \node[below of=HO-22,node distance=0.2 cm] {$1-\epsilon$};
\end{tikzpicture}

\caption{A composite fading and additive noise channel with binary alphabets. We consider an ON-OFF fading channel with ON probability $p$ and an additive noise channel with symmetric transition probabilities $\epsilon$.}
\label{fig2_fading}
\end{center}
\end{figure}

\begin{remarks}
\end{remarks}
\begin{enumerate}
\item We note that the sync frame length needed for frame synchronisation increases with channel fading
(with OFF probability $1-p$).
\item In Figure~\ref{fig2_fading}, we have illustrated a composite ON-OFF fading and additive noise channel
with binary alphabets. The transition probabilities $Q_n(\cdot|\cdot)$ for the
additive noise channel is
\[ Q_n = \left[ \begin{array}{ll} 1 - \epsilon & \epsilon \\ \epsilon & 1-\epsilon \end{array} \right] \]
and the corresponding synchronisation threshold $\alpha(Q_n)$ is
\[ \alpha(Q_n) = (1-\epsilon) \log\left(\frac{1-\epsilon}{\epsilon} \right) + \epsilon \log\left( \frac{\epsilon}{1 - \epsilon} \right) \]
The transition probabilities $Q(\cdot|\cdot)$ for the composite channel is
\[ Q = \left[ \begin{array}{cc} 1 - \epsilon & \epsilon \\ p \epsilon + (1-p)(1-\epsilon) & p (1-\epsilon) + (1-p) \epsilon \end{array} \right] \]
and the synchronisation threshold $\alpha(Q)$ is given by
\[ \alpha(Q) = (1-\epsilon_p) \log\left(\frac{1-\epsilon_p}{\epsilon} \right) + \epsilon_p \log\left( \frac{\epsilon_p}{1 - \epsilon} \right) \]
where $\epsilon_p = (1 - p)(1 - \epsilon) + p \epsilon$.
Taking $\epsilon \rightarrow 0$, we see that,
\begin{eqnarray*}
 Q \approx \left[ \begin{array}{ll} 1 - \epsilon & \epsilon \\ 1-p & p \end{array} \right]  
 \quad \text{and} \quad 
 \lim_{\epsilon \rightarrow 0} \frac{\alpha(Q)}{\alpha(Q_n)} = p 
\end{eqnarray*}
i.e., the bound in Lemma~\ref{lem:fading} is tight.
\end{enumerate}

\section{Rayleigh Fading and AWGN channel}
\label{sec:rayleigh_fading}
\ifnum\DEFARXIV=1
	\begin{figure}
	\begin{center}
	\def\layersep{2.5cm}
	\begin{tikzpicture}[draw=black!50, node distance=\layersep,font=\scriptsize]
	    \tikzstyle{sym}=[circle,fill=black!20,minimum size=5pt,inner sep=0pt]
	    \tikzstyle{xsym}=[sym, color=black!60,text=black];
	    \tikzstyle{ysym}=[sym, color=black!40,text=black];
	    \tikzstyle{dysym}=[sym, color=black!40,opacity=0.5,text=black,minimum size=4pt];
	    \tikzstyle{hsym}=[sym, color=black!30,text=black];
	    \tikzstyle{dhsym}=[sym, color=black!30, opacity=0.5,text=black,minimum size=4pt];
	    \tikzstyle{annot} = [text width=4em, text centered]
	
	    \foreach \name / \y in {1,2}
	    {
	    	\pgfmathsetmacro{\yn}{-2*\y}%
			\path[yshift=0.5cm]
				node[xsym] (I-\name) at (0,\yn) {};
	        \node[left of=I-\name,node distance=10] {$x_\y$};
	     }

	    \foreach \name / \y in {1,2,3}
	    {
	    	\pgfmathsetmacro{\yn}{-2*\y+1}%
	        \path[yshift=0.5cm]
	            node[hsym] (H-\name) at (\layersep,\yn cm) {};
	    }
	    \foreach \y in {1,...,19}
	    {
	    	\pgfmathsetmacro{\yn}{-1-\y/5}%
	        \path[yshift=0.5cm]
	            node[dhsym] at (\layersep,\yn cm) {};
	    }

	    \foreach \name / \y in {1,2}
	    {
	    	\pgfmathsetmacro{\yn}{-2*\y}%
		 \path[yshift=0.5cm]
		  node[ysym] (O-\name) at (2*\layersep,\yn) {};
		 \node[right of=O-\name,node distance=10] {$y_\y$};
		}
	    \foreach \y in {1,...,9}
	    {
	    	\pgfmathsetmacro{\yn}{-2-2*\y/10}%
	        \path[yshift=0.5cm]
	            node[dysym] at (2*\layersep,\yn cm) {};
	    }

	    \foreach \source in {1,...,2}
	        \foreach \dest in {1}
	            \path [line width=1pt] (I-\source) edge node (IH-\source\dest) {} (H-\dest);
	    \foreach \source in {1,...,2}
	        \foreach \dest in {2,3}
	            \path[dashed] (I-\source) edge node (IH-\source\dest) {} (H-\dest);
	
	    \foreach \source in {1}
	        \foreach \dest in {1}
	            \path [line width=1pt] (H-\source) edge node (HO-\source\dest) {} (O-\dest);
	    \foreach \source in {1,2,3}
	        \foreach \dest in {1,...,2}
	            \path[dashed] (H-\source) edge node (HO-\source\dest) {} (O-\dest);
	
	    \node[annot,above of=H-1, node distance=1cm] (hl) {$\mathcal{H}$};
	    \node[annot,left of=hl] {$\mathcal{X}$};
	    \node[annot,right of=hl] {$\mathcal{Y}$};
	    \node[above of=HO-11,node distance=0.4 cm] {$P(y_1|h_1)$};
	    \node[above of=IH-11,node distance=0.4 cm] {$P(h_1|x_1)$};
	\end{tikzpicture}
	\caption{A representation of Rayleigh fading and additive noise channel with infinite alphabets for $\mathcal{H}$ and $\mathcal{Y}$. }
	\label{fig:rayleigh_channel}
	\end{center}
	\end{figure}
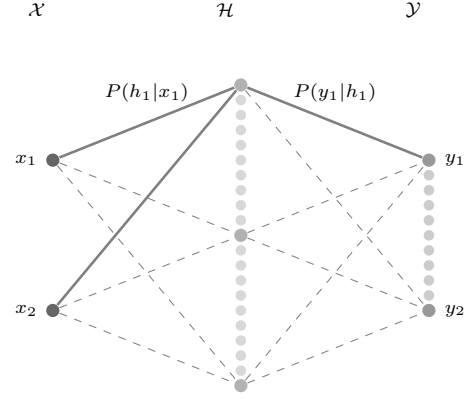
\fi 

In this section, we characterise the synchronisation threshold
for the Rayleigh fading and AWGN channel. 
We make the following assumptions about the channel.
\begin{enumerate}
\item The received signal $y_n$ in slot $n$ is modelled as $y_n = h_n x_n + n_n$ 
for all $n$, where $h_n$ is Rayleigh (with scale parameter $\sigma_H$)
and $n_n$ is WGN with variance $\sigma^2$.
\item We assume that the Rayleigh fading
and the additive Gaussian noise are independent over slots and independent of each other as well.
\item We consider a binary input alphabet set $\{ x(0) = 0, x(1) = \sqrt{P} \}$ for the composite
channel. $P$ could correspond to the symbol power constraint and $\frac{P}{\sigma^2}$ would then be the SNR.
We note that it is sufficient to consider the binary alphabet set for the sequential frame
synchronisation problem (see Section~\ref{sec:setup} or \cite{Chandar2008} for details).
\item We consider a continuous alphabet set $(-\infty,\infty)$ for the wireless channel (due to the
Rayleigh fading and the additive Gaussian noise).
The framework developed in the previous section is limited to channel with a finite alphabet set. We have
used a large but finite alphabet set in our simulations and have used a limiting approximation to obtain closed-form expressions for the synchronisation threshold.
\end{enumerate}
From equation~(\ref{eqn:alpha}), we know that 
\begin{eqnarray*}
\alpha(Q) &=& \underset{x\in\mathcal{X}}{\max\,} D(Q(\cdot|x) \| Q(\cdot|x(0)))\\
&=& D(Q(\cdot|x(1)) \| Q(\cdot|x(0)))
\end{eqnarray*}
where $x(1)= \sqrt{P}$ and $x(0)=0$. The conditional density functions that characterizes the
channel transition probabilities are
\[Q(y|x(0)) = {\mathsf P}(n=y) = \frac{1}{\sqrt{2\pi \sigma^2}}e^{-\frac{y^2}{2\sigma^2}}, -\infty<y<\infty\]
and
\begin{IEEEeqnarray*}{rCl}
 Q(y|x(1)) &=& {\mathsf P}(h\sqrt{P}+n=y)\\
 &=& \int_0^\infty  \frac{ e^{-\frac{(y-h\sqrt{P})^2}{2\sigma^2}}
 						}{\sqrt{2\pi \sigma^2}} 
		    \frac{h e^{-\frac{h}{2\sigma_H^2}}
		    	  }{\sigma_H^2}  dh, -\infty < y < \infty \\
\end{IEEEeqnarray*}
For the continuous alphabet set, we can approximate $\alpha(Q)$ as
\[ \alpha(Q) = \int_{-\infty}^\infty Q(y|x_1) \log \frac{Q(y|x_1)}{Q(y|x_0)} dy\]
We will now characterize $\alpha(Q)$ through numerical evaluation.
\begin{remarks}
\end{remarks}
\begin{enumerate}
\item In Figure~\ref{fig:alpha_Rayleigh_ratio}, we plot the ratio of the synchronisation threshold 
of the composite channel $\alpha(Q)$ and the synchronisation threshold of the AWGN
channel $\alpha(Q_n) = \frac{P}{2 \sigma^2}$ (see \cite{Chandar2008}) 
as a function of the SNR ($\frac{P}{\sigma^2}$) and for different values
of $\sigma_H$. From the figure, we note that $\alpha(Q)$ is linear with SNR for large SNR,
i.e., $\alpha(Q) \propto \frac{P}{\sigma^2}$ and hence, $\alpha(Q) \propto \alpha(Q_n)$ as well.
Also, from Figure~\ref{fig:alpha_Rayleigh_ratio} (and numerical verification), we observe that,
\begin{equation}
\alpha(Q) \simeq \alpha(Q_n) \cdot 2 \sigma_H^2 \label{eqn:rayleigh_approx}
\end{equation}
\item We note that the synchronisation threshold $\alpha(Q)$ of the Rayleigh channel can be strictly greater than $\alpha(Q_n)$ 
(of the AWGN channel) for large values of the Rayleigh scale parameter $\sigma_H$. Unlike the ON-OFF fading channel
discussed in Section~\ref{sec:fading}, the output of the Rayleigh fading channel can have a gain larger than $1$ depending 
on the channel realisation. Equation~(\ref{eqn:rayleigh_approx}) captures the effect of both the fading channel gain and the 
channel transition probabilities.
\end{enumerate}

\ifnum\DEFARXIV=1
	\begin{figure}
	\begin{center}
	\begin{tikzpicture}
	\pgfplotstableread{plot_alpha_mathematica.txt} \data
	\begin{axis}
	[
	    legend style={at={(0.02,0.98)},anchor=north west},
	    y label style={rotate=270},
	    xlabel={SNR (linear)},
	    ylabel={$\alpha(Q)$},
	    ymajorgrids=true,
	    grid style=dashed,
	    font=\footnotesize,
	 	tick scale binop=\mathsf{x}\,,
	    cycle list name=mark list,
	]
	    \addplot+[color=red] table[x=1, y=4] \data;
	    \addplot+[color=blue] table[x=1, y=5] \data;
	    \addplot+[color=brown] table[x=1, y=6] \data;
	    \addlegendentry{$\sigma_H=1$};
	    \addlegendentry{$\sigma_H=2$};
	    \addlegendentry{$\sigma_H=3$};
	\end{axis};
	
	 \end{tikzpicture}
	\end{center}
	\caption{ Plot of synchronisation threshold of the Rayleigh fading channel, $\alpha(Q)$, for different values of $\sigma_H$. }
	\label{fig:alpha_Rayleigh}
	\end{figure}
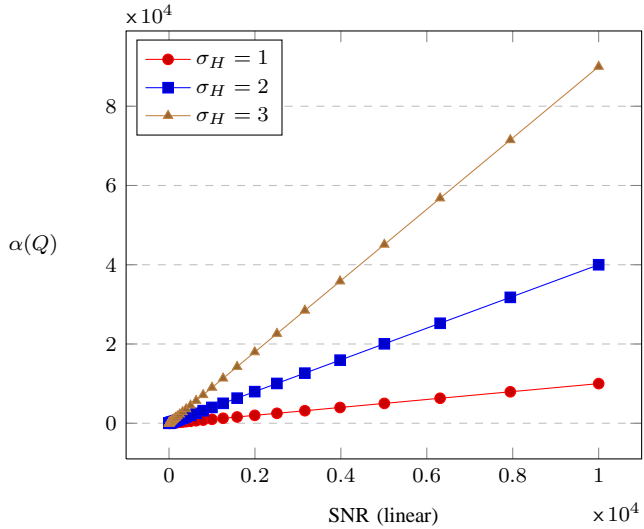
\fi 

\begin{figure}
\begin{center}
\begin{tikzpicture}
\pgfplotstableread{plot_alpha_mathematica.txt} \data
\begin{axis}
[
    legend style={at={(0.98,0.7)},anchor=east},
    y label style={rotate=270},
    xlabel={SNR (linear) },
    ylabel={$\frac{\alpha(Q)}{\alpha(Q_n)}$},
    extra y ticks={2,8,18},
    ymajorgrids=true,
    grid style=dashed,
    font=\footnotesize,
 	tick scale binop=\mathsf{x}\,,
    cycle list name=mark list,
]
    \addplot+[color=red] table[x=1, y=9] \data;
    \addplot+[color=blue] table[x=1, y=10] \data;
    \addplot+[color=brown] table[x=1, y=11] \data;
    \addlegendentry{$\sigma_H=1$};
    \addlegendentry{$\sigma_H=2$};
    \addlegendentry{$\sigma_H=3$};
\end{axis};
\end{tikzpicture}
\end{center}
\caption{ Ratio of synchronisation thresholds  of the Rayleigh fading channel, $\alpha(Q)$, and the AWGN channel, $\alpha(Q_n)$, vs SNR, $\frac{P}{\sigma^2}$. }
\label{fig:alpha_Rayleigh_ratio}
\end{figure}
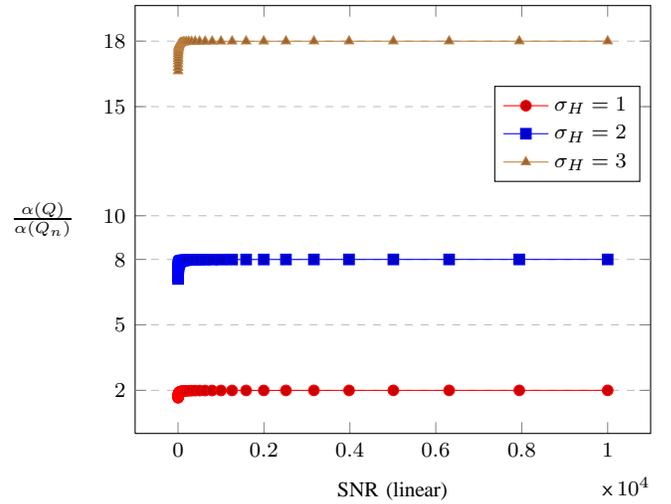

\section{A General Framework for Asynchronism}
\label{sec:generalization}
In \cite{Chandar2008}, \cite{Tchamkerten2009} and even in the previous sections of this paper,
we studied synchronisation threshold for the channel that would characterise the 
sync frame length $N$ needed for error-free frame synchronization.
In this section, we will propose a framework that permits a tradeoff between sync frame length $N$ and
channel $Q$ to support asynchronism. The framework will allow us to characterise the scaling needed
of the sync packet energy for asymptotic error-free frame synchronisation.

Consider a sequence of triples, channel, sync word and sequential decoder,
$( \{ {\mathcal X}_A, {\mathcal Y}_A, Q_A \}, {\mathsf s}^{N_A}, \hat{v} )$
defined for all $A \geq 1$, where $A$ is the asynchronous interval length.
Define $\alpha(Q_A)$ as
\[ \alpha(Q_A) = \max_{x \in {\mathcal X}_A} D( Q_A(\cdot|x) \| Q_A(\cdot|x(0)) ) \]
The following theorem generalizes Theorem~1
in \cite{Chandar2008} and discusses
the necessary scaling needed of $N_A$ and $\alpha(Q_A)$ for asymptotic error-free frame synchronisation.
\begin{theorem}
\label{thm:generalization}
Consider a sequence of triples, $( \{ {\mathcal X}_A, {\mathcal Y}_A, Q_A \}, {\mathbf s}^{N_A}, \hat{v} )$ parameterized
by the period $A$. Let $N_A \rightarrow \infty$ as $A \rightarrow \infty$ and let $\alpha(Q_A)$ be non-decreasing in $A$.
Then, the probability of frame detection error ${\mathsf {P_A}}(\{\hat{v} \neq v\}) \rightarrow 0$ if
$e^{\alpha(Q_A) N_A} > A$. \qed
\end{theorem}

\begin{remarks}
\end{remarks}
\begin{enumerate}
\item Theorem~\ref{thm:generalization} characterizes the rate at which $\alpha(Q_A) \times N_A$ must scale with $A$
for the frame detection error to tend to zero (asymptotically). In \cite{Chandar2008}, the channel was assumed
to be the same independent of $N$ or $A$. The generalisation proposed in Theorem~\ref{thm:generalization}
enables us to study the tradeoff between $\alpha(Q_A)$ and $N_A$ for supporting asynchronism.
\item For an AWGN channel, we know that $\alpha(Q_n) = \frac{P}{2 \sigma^2}$.
Hence, $\alpha(Q_A) N_A \propto P_A N_A$ (the energy of the sync packet).
Thus, the above theorem also characterizes the necessary scaling needed of the energy of the sync packet
for the frame detection error to tend to zero.
\end{enumerate}

\ifnum\DEFARXIV=1
	Here, we have presented only a necessary outline of the proof for Theorem~\ref{thm:generalization} as the argument
	is similar to the presentation in \cite{Chandar2008}.
	\begin{proof}
	We consider the framework presented in Section~\ref{sec:setup} for every $A$.
	A sync packet $\mathbf{s}^{N_A}$ of length $N_A$ is transmitted at some random time $v \sim U\{1,A\}$.
	The discrete memory-less channel is characterised by finite input and output alphabet sets
	${\mathcal X}_A$ and ${\mathcal Y}_A$ respectively, and transition probabilities $Q_A(\cdot|\cdot)$
	with
	\[ \alpha(Q_A) = \max_{x \in {\mathcal X}_A} D( Q_A(\cdot|x) \| Q_A(\cdot|x(0)) ) \]
	
	Following \cite{Chandar2008}, we consider a sync sequence ${\mathbf s}^{N_A}$ of length $N_A$ with the following
	properties.
	\begin{enumerate}
	\item Fix some large $K$, where $K$ is any integer such that $\lfloor{\frac{N_A}{K}} \rfloor=2^m - 1$ for some $m = 1,2,\cdots$.
	Let $s_n = x(1)$ for $\lfloor \frac{N_A}{K} \rfloor < n \leq N_A$. Consider a maximal-length
	shift register sequence (MLSR) $\{m_n : n = 1,2,\cdots,\lfloor \frac{N_A}{K} \rfloor\}$
	of length $\lfloor \frac{N_A}{K} \rfloor$ and map it to $\{ s_n : n = 1,2,\cdots,
	\lfloor \frac{N_A}{K} \rfloor \}$ such that $s_n = x(1)$ if $m_n = 0$ and $s_n = x(0)$ if $m_n = 1$.
	 \item The Hamming distance between the sync sequence ${\mathbf s}^{N_A}$ and
	 any of its shifted sequences is now $\Omega(N_A)$.
	\end{enumerate}
	
	We consider a sequential joint typicality decoder for the problem setup. At every time $t+N_A-1$,
	the decoder computes the empirical joint distribution $\hat{\mathsf P}$ induced by the sync pattern
	${\mathbf s}^{N_A}$ and the previous $N_A$ output symbols $\{ y_{t}, \cdots, y_{t+N_A-1}\}$. 
	\begin{equation*}
	\mathsf{\hat{P}}_{\mathbf{s},\mathbf{y}}(x,y) = \frac{\mathsf{N}(x,y)}{N_A}, \text{ for all } (x,y)\in \mathcal{X} \mathsf{x} \mathcal{Y}
	\end{equation*}
	where, $\mathsf{N}(x,y)$ denotes the number of joint occurrences of $(x,y)$ in the sync code word and the channel output.
	The expected joint distribution, ${\mathsf P}$, induced by the sync pattern on the channel output, is defined as 
	\begin{equation*}
	\mathsf{P}_{\mathbf{s},\mathbf{y}}(x,y) \triangleq \mathsf{\hat{P}}_{\mathbf{s}}(x) Q(y|x)
	\end{equation*}
	where, $\mathsf{\hat{P}}_{\mathbf{s}}(x) = \frac{\mathsf{N}(x)}{N_A}, \text{ for all } x \in \mathcal{X}$ with $\mathsf{N}(x)$ denoting the number of occurrences of $x$ in the sync code word. 
	If the
	empirical distribution is close enough to the expected joint distribution , i.e., if $| \hat{\mathsf P} - {\mathsf P}| \leq \mu$ (for some $\mu > 0$), then, the decoder
	stops and declares $\hat{v} = t$.
	\begin{figure}
	\begin{center}
	\begin{tikzpicture}[scale=1,font=\scriptsize]
	\begin{axis}
	[	
	    axis y line=none,
	    axis x line=none,    
	    ymin=-1.2, ymax=3.2,
	    height = 4 cm, width = 5.5 cm,    
		title style={at={(axis cs:\xmid,-2.2)}},title=a. Error-less,font=\footnotesize,    
	]
		\def \xend{100}
		\def \xmid{50}
		\def \xbeg{1}
		\def \pbeg{71}
		\def \pend{80}
	    \addplot[ycomb,black!20,samples=\pbeg-1-\xbeg,domain=\xbeg:\pbeg-1] {rand};
	    \addplot[ycomb,black!70,samples=\pend-\pbeg,domain=\pbeg:\pend] {2+rand};
	    \addplot[ycomb,black!20,samples=\xend-\pend-1,domain=\pend+1:\xend] {rand};
	   	\pgfmathsetmacro{\ya}{\pbeg-.2}
	   	\pgfmathsetmacro{\yb}{\pend+.2}
	    \draw [black!70] (axis cs:\ya,-0.1) rectangle (axis cs:\yb,3);
	   	\pgfmathsetmacro{\yn}{\pbeg-1}
	    \node[below left , black,align=center,font=\scriptsize] at (axis cs:\yn,3) {Sync Packet};
	 \end{axis};
	 \end{tikzpicture}
	\begin{tikzpicture}[scale=1,font=\scriptsize]
	\begin{axis}
	[	
	    axis y line=none,
	    axis x line=none,    
	    ymin=-1.2, ymax=3.2,
	    height = 4 cm, width = 5.5 cm,    
		title style={at={(axis cs:\xmid,-2.2)}},title=b. Error E1,font=\footnotesize,    
	]
		\def \xend{100}
		\def \xmid{50}
		\def \xbeg{1}
		\def \pbeg{71}
		\def \pend{80}
		\def \nbeg{31}
		\def \nend{40}
	    \addplot[ycomb,black!20,samples=\nbeg-1-\xbeg,domain=\xbeg:\nbeg-1] {rand};
	    \addplot[ycomb,black!40,samples=\nend-\nbeg,domain=\nbeg:\nend] {1+rand};
	    \addplot[ycomb,black!20,samples=\pbeg-1-\nend-1,domain=\nend+1:\pbeg-1] {rand};
	    \addplot[ycomb,black!70,samples=\pend-\pbeg,domain=\pbeg:\pend] {2+rand};
	    \addplot[ycomb,black!20,samples=\xend-\pend-1,domain=\pend+1:\xend] {rand};
	   	\pgfmathsetmacro{\ya}{\pbeg-.2}
	   	\pgfmathsetmacro{\yb}{\pend+.2}
	    \draw [black!70] (axis cs:\ya,0.1) rectangle (axis cs:\yb,3);
	   	\pgfmathsetmacro{\na}{\nbeg-.2}
	   	\pgfmathsetmacro{\nb}{\nend+.2}
	    \draw [black!30] (axis cs:\na,0.1) rectangle (axis cs:\nb,3);
	   	\pgfmathsetmacro{\nc}{\nbeg-1}
	    \node[below left , black,align=center,font=\scriptsize]  at (axis cs:\nc,3)
		    {Sync-like\\noise};
	 \end{axis};
	 \end{tikzpicture}
	\begin{tikzpicture}[scale=1,font=\scriptsize]
	\begin{axis}
	[	
	    axis y line=none,
	    axis x line=none,    
	    ymin=-1.2, ymax=3.2,
	    height = 4 cm, width = 5.5 cm,    
	 title style={at={(axis cs:\xmid,-2.2)}},title=c. Error E2,font=\footnotesize,   
	]
		\def \xend{100}
		\def \xmid{50}
		\def \xbeg{1}
		\def \pbeg{71}
		\def \pend{80}
		\def \nbeg{65}
		\def \nend{74}
	    \addplot[ycomb,black!20,samples=\nbeg-1-\xbeg,domain=\xbeg:\nbeg-1] {rand};
	    \addplot[ycomb,black!40,fill=black!30,samples=\pbeg-1-\nbeg,domain=\nbeg:\pbeg-1] {1+rand};
	    \addplot[ycomb,black,fill=black!20,samples=\nend-\pbeg,domain=\pbeg:\nend] {2+rand};
	    \addplot[ycomb,black!70,fill=black!70,samples=\pend-\nend-1,domain=\nend+1:\pend] {2+rand};
	    \addplot[ycomb,black!20,samples=\xend-\pend-1,domain=\pend+1:\xend] {rand};
	   	\pgfmathsetmacro{\ya}{\pbeg-.2}
	   	\pgfmathsetmacro{\yb}{\pend+.2}
	    \draw [black!70] (axis cs:\ya,-0.2) rectangle (axis cs:\yb,3.1);
	   	\pgfmathsetmacro{\na}{\nbeg-.2}
	   	\pgfmathsetmacro{\nb}{\nend+.2}
	    \draw [black!30] (axis cs:\na,-0.1) rectangle (axis cs:\nb,3);
	   	\pgfmathsetmacro{\nc}{\nbeg-1}
	   	\node[below left , black,font=\scriptsize,align=center]  at (axis cs:\nc,3)
	    	{Sync-like\\Overlap};
	 \end{axis};
	 \end{tikzpicture}
	\begin{tikzpicture}[scale=1,font=\scriptsize]
	\begin{axis}
	[	
	    axis y line=none,
	    axis x line=none,    
	    ymin=-1.2, ymax=3.2,
	    height = 4 cm, width = 5.5 cm,    
	title style={at={(axis cs:\xmid,-2.2)}},title=d. Error E3,font=\footnotesize,    
	]
		\def \xend{100}
		\def \xmid{50}
		\def \xbeg{1}
		\def \pbeg{71}
		\def \pend{80}
	    \addplot[ycomb,black!20,samples=\pbeg-1-\xbeg,domain=\xbeg:\pbeg-1] {rand};
	    \addplot[ycomb,black!40,samples=\pend-\pbeg,domain=\pbeg:\pend] {1+3*rand};
	    \addplot[ycomb,black!20,samples=\xend-\pend-1,domain=\pend+1:\xend] {rand};
	   	\pgfmathsetmacro{\ya}{\pbeg-.2}
	   	\pgfmathsetmacro{\yb}{\pend+.2}
	    \draw [black!30] (axis cs:\ya,-1.1) rectangle (axis cs:\yb,3.1); 
	   	\pgfmathsetmacro{\ya}{\pbeg-1}
	    \node[below left , black,align=center,font=\scriptsize] at (axis cs:\ya,3.1)
	      {Noise-like \\sync};
	 \end{axis};
	 \end{tikzpicture}
	\end{center}
	\caption{ Examples for each partition of the error event $\{\hat{a} \neq a\}$ during packet reception}
	\label{fig:pkt_errors}
	\end{figure}
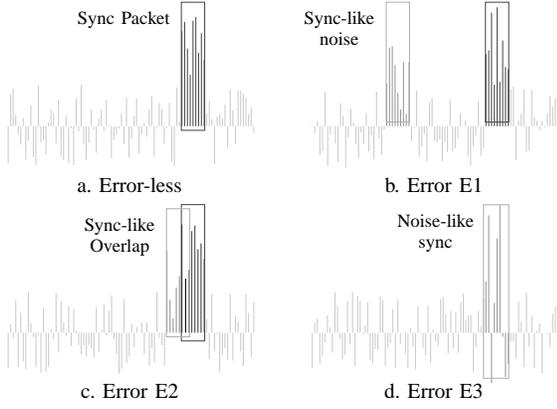
	
	The error event $\{ \hat{v} \neq v \}$ can be partitioned as discussed below (and as illustrated
	in the Figure~\ref{fig:pkt_errors}).
	\begin{itemize}
	    \item $E_1 : \hat{v} \in  \{1, \cdots, v-N\} \cup \{v+1, \cdots, A\}$. This corresponds to the event that the output symbols generated entirely by the zero input $x(0)$ is jointly typical.
	    \item $E_2 : \hat{v} \in \{v-N+1, \cdots, v-1\}$. This corresponds to the event that the output symbols generated partially by $x(0)$ and sync word is jointly typical.
		\item $E_3 : \hat{v} \notin \{ v \}$. This corresponds to the event that the output symbols generated by the sync word is not jointly typical. 
	\end{itemize}
	We note that the event $E_1 \cup E_2$ does not contain the event $E_3$ as we consider sequential
	frame detection.
	Using a union bound, we get,
	\[ {\mathsf P}(\{ \hat{v} \neq v \}) \leq {\mathsf P}(E_1) + {\mathsf P}(E_2) + {\mathsf P}(E_3) \]
	We will now show that $P(E_1), P(E_2)$ and $P(E_3)$ tend to zero for the listed conditions.
	Suppose that $A < e^{N_A \alpha(Q_A)}$ for all $A$. 
	Now, consider an asynchronism length of $\bar{A}=e^{N_A \epsilon_1 (\alpha(Q_A) - \epsilon_2)}$, where
	$0 < \epsilon_1 < 1$ and $\epsilon_2 > 0$. Clearly, $\bar{A}  < e^{N_A \alpha(Q_A)}$. 
	
	The probability that the channel output for an
	input sequence composed entirely of $x(0)$ is jointly typical can be computed as discussed in \cite{Chandar2008},
	\[ e^{-N_A(1-\frac{1}{K}) [ D( Q_A(\cdot|x(1)) || Q_A(\cdot|x(0)) ) + H(Q_A(\cdot|x(1))) - \delta]} \]
	where $\delta$ is a function of $\mu$ (of the joint typicality decoder) and tends to zero as $\mu \rightarrow 0$.
	Substituting for $D( Q_A(\cdot|x(1)) || Q_A (\cdot|x(0)) ) = \alpha(Q_A)$ in the above expression and ignoring the non-negative entropy term, we get
	the following upper bound
	\[ e^{-N_A(1-\frac{1}{K}) [ \alpha(Q_A) - \delta]} \]
	We can now bound ${\mathsf P}(E_1)$ using a union bound as,
	\begin{eqnarray*}
	{\mathsf P}(E_1) &\leq& \bar{A} \times e^{-N_A (1-\frac{1}{K}) (\alpha(Q_A) - \delta)}\\
	 &=& e^{N_A \epsilon_1(\alpha(Q_A) - \epsilon_2) }e^{-N_A(1-\frac{1}{K}) (\alpha(Q_A) - \delta)}\\
	&=& e^{N_A \alpha(Q_A) (\epsilon_1 - (1 - \frac{1}{K}))} e^{-N_A (\epsilon_1 \epsilon_2 - \delta)}
	\end{eqnarray*}
	For large $K$, small $\delta$ (depends on $\mu$ of the typicality decoder) and as $A \rightarrow \infty$ (and hence,
	$N_A \alpha(Q_A) \rightarrow \infty$), we have, ${\mathsf P}(E_1) \rightarrow 0$ for any $\epsilon_1$ and $\epsilon_2$.
	Thus $\mathsf{P}(E_1) \rightarrow 0$ for all $A < e^{N_A \alpha(Q_A)}$
	
	The probability that the channel output for an
	input sequence composed partially of $x(0)$ and $x(1)$ (a shifted sequence) is jointly typical can be upper bounded as
	\begin{eqnarray*}
	&e^{-N_A \frac{1}{K} [ D( Q(\cdot|x(1)) || Q (\cdot|x(0)) ) + H(Q(\cdot|x(1))) - \delta]}&\\
	&\leq  e^{-N_A \frac{1}{K} [ \alpha(Q_A) - \delta]}&
	\end{eqnarray*} 
	Using a union bound for ${\mathsf P}(E_2)$, we get,
	\[ {\mathsf P}(E_2) \leq N_A e^{-N_A \frac{1}{K} [ \alpha(Q_A) - \delta]} \]
	For small $\delta$ and as $A \rightarrow \infty$ (and hence, $N_A \alpha(Q_A) \rightarrow \infty$), we have ${\mathsf P}(E_2) \rightarrow 0$.
	
	If $N_A \rightarrow \infty$ as $A \rightarrow \infty$, then ${\mathsf P}(E_3) \rightarrow 0$ (follows from the
	weak law of large numbers).
	
	Thus, we have ${\mathsf P}(\{\hat{v} \neq v\}) \rightarrow 0$ if $e^{N_A \alpha(Q_A)} > A$.
	The above arguments show the achievability of asymptotic error-free frame synchronisation
	if $e^{\alpha(Q_A) N_A} > A$.
	The converse follows directly from the discussion in \cite{Chandar2008}, if we consider the special case of $\alpha(Q_A) = \alpha(Q)$ and 
	let $N_A \rightarrow \infty$.
	\end{proof}
\fi 

The following corollary discusses the application of the Theorem~\ref{thm:generalization} to an AWGN channel.
\begin{corollary}
\label{cor:energy}
Consider an AWGN channel with symbol power $P_A$ and noise variance $\sigma^2$.
Let $N_A \rightarrow \infty$ as $A \rightarrow \infty$ and let $P_A$ be non-decreasing in $A$.
Then, the probability of frame detection error ${\mathsf {P_A}}(\{\hat{v} \neq v\}) \rightarrow 0$
if $e^{\frac{1}{2 \sigma^2} N_A P_A} > A$. \qed
\end{corollary}
\ifnum\DEFARXIV=1
	\begin{proof}
	From \cite{Chandar2008}, we know that $\alpha(Q_A) = \frac{P_A}{2 \sigma^2}$ for an AWGN channel.
	From Theorem~\ref{thm:generalization}, we have ${\mathsf {P_A}}(\{\hat{v} \neq v\}) \rightarrow 0$
	if $e^{\alpha(Q_A) N_A} = e^{\frac{P_A}{2\sigma^2}N_A}> A$.
\end{proof}
\fi 
\begin{remarks}
\end{remarks}
\begin{enumerate}
\item Define $E_A = N_A \times P_A$ as the energy of the sync packet. Then, the above corollary characterises
the scaling necessary of the energy of the sync packet for asymptotic error-free frame synchronisation.
We note that the synchronisation threshold for the AWGN channel with respect to the sync packet energy
is $\frac{1}{2 \sigma^2}$.
\item A similar result holds for the Rayleigh fading and AWGN channel. The synchronization threshold
with respect to the sync packet energy for the composite channel would be 
$\frac{\sigma^2_H}{\sigma^2}$ (see Figure~\ref{fig:alpha_Rayleigh_ratio}).
\end{enumerate}

\section{Conclusion}
In this paper, we have studied a sequential frame synchronization problem
for a fading channel and additive noise.
For an ON-OFF fading channel with ON probability $p$ and an additive noise channel with
transition probabilities $Q_n$, we characterized the synchronization threshold of the composite
channel $\alpha(Q)$
and showed that $\alpha(Q) \leq \alpha(Q_n) \, p$.
For a Rayleigh fading and AWGN channel, we characterised the synchronisation threshold
as $\alpha(Q) \simeq \alpha(Q_n) \, 2\sigma_H^2$, where $\sigma_H$ is the scale parameter 
of the Rayleigh channel.
Finally, we proposed a framework that permits a trade-off between sync word length $N$ and
channel $Q$ to support asynchronism.
The framework allowed us to characterise the synchronisation threshold for AWGN channel
in terms of the sync frame energy (i.e., $e^{\frac{1}{2 \sigma^2} E} > A$) instead of the sync frame length $N$.

The sequential frame synchronisation problem is related to the quickest transient change detection problems studied 
in works such as \cite{premkumar2010}. In the future, we seek to generalize the frame synchronisation framework
with general definitions for frame synchronisation and the error events.

\bibliographystyle{IEEEtran}
\bibliography{bibfile_short}
\fi 

\ifnum\DEFAPPENDIX=1
\pagebreak
\section*{Appendix}
\subsection{Proof of Lemma~\ref{lem:fading}}
	\begin{proof}
	Define $x(1) := \underset{x \in {\mathcal X}}{\arg\,\max\,} D( Q(\cdot|x) \| Q(\cdot|x(0)) )$.
	Then,
	\begin{IEEEeqnarray}{rCl}
	\alpha(Q) &=& \max_{x \in {\mathcal X}} D( Q(\cdot|x) \| Q(\cdot|x(0)) ) \nonumber \\
	&=& D( Q(\cdot|x(1)) \| Q(\cdot|x(0)) ) \nonumber \\
	&=& D( p \ Q_n(\cdot|x(1)) + (1-p) \ Q_n(\cdot|x(0)) 
			\| Q_n(\cdot|x(0)) ) \nonumber \\
	&\leq& p \ D(Q_n(\cdot|x(1)) \| Q_n(\cdot|x(0)) \label{eqn:fading}
	\end{IEEEeqnarray}
	The last equation follows from Jensen's inequality. Also, we know that
	for an independent fading and additive noise model,
	\begin{IEEEeqnarray*}{rCl}
	\IEEEeqnarraymulticol{3}{l}{\arg\max_{x \in {\mathcal X}} D( Q(\cdot|x) \| Q(\cdot|x(0)) )} \\
	&=& x(1) \\
	&=& \arg\max_{x \in {\mathcal X}} D( Q_n(\cdot|x) \| Q_n(\cdot|x(0)) )
	\end{IEEEeqnarray*}
	Thus, we have, $D(Q_n(\cdot|x(1)) \| Q_n(\cdot|x(0)) = \alpha(Q_n)$ and, substituting in Equation~(\ref{eqn:fading}), we get,
	\[ \alpha(Q) \leq p \ \alpha(Q_n) \]
	\end{proof}
\subsection{Proof of Theorem~\ref{thm:generalization}}
	Here, we have presented only a necessary outline of the proof for Theorem~\ref{thm:generalization} as the argument
	is similar to the presentation in \cite{Chandar2008}.
	\begin{proof}
	We consider the framework presented in Section~\ref{sec:setup} for every $A$.
	A sync packet $\mathbf{s}^{N_A}$ of length $N_A$ is transmitted at some random time $v \sim U\{1,A\}$.
	The discrete memory-less channel is characterised by finite input and output alphabet sets
	${\mathcal X}_A$ and ${\mathcal Y}_A$ respectively, and transition probabilities $Q_A(\cdot|\cdot)$
	with
	\[ \alpha(Q_A) = \max_{x \in {\mathcal X}_A} D( Q_A(\cdot|x) \| Q_A(\cdot|x(0)) ) \]
	
	Following \cite{Chandar2008}, we consider a sync sequence ${\mathbf s}^{N_A}$ of length $N_A$ with the following
	properties.
	\begin{enumerate}
	\item Fix some large $K$, where $K$ is any integer such that $\lfloor{\frac{N_A}{K}} \rfloor=2^m - 1$ for some $m = 1,2,\cdots$.
	Let $s_n = x(1)$ for $\lfloor \frac{N_A}{K} \rfloor < n \leq N_A$. Consider a maximal-length
	shift register sequence (MLSR) $\{m_n : n = 1,2,\cdots,\lfloor \frac{N_A}{K} \rfloor\}$
	of length $\lfloor \frac{N_A}{K} \rfloor$ and map it to $\{ s_n : n = 1,2,\cdots,
	\lfloor \frac{N_A}{K} \rfloor \}$ such that $s_n = x(1)$ if $m_n = 0$ and $s_n = x(0)$ if $m_n = 1$.
	 \item The Hamming distance between the sync sequence ${\mathbf s}^{N_A}$ and
	 any of its shifted sequences is now $\Omega(N_A)$.
	\end{enumerate}
	
	We consider a sequential joint typicality decoder for the problem setup. At every time $t+N_A-1$,
	the decoder computes the empirical joint distribution $\hat{\mathsf P}$ induced by the sync pattern
	${\mathbf s}^{N_A}$ and the previous $N_A$ output symbols $\{ y_{t}, \cdots, y_{t+N_A-1}\}$. 
	\begin{equation*}
	\mathsf{\hat{P}}_{\mathbf{s},\mathbf{y}}(x,y) = \frac{\mathsf{N}(x,y)}{N_A}, \text{ for all } (x,y)\in \mathcal{X} \mathsf{x} \mathcal{Y}
	\end{equation*}
	where, $\mathsf{N}(x,y)$ denotes the number of joint occurrences of $(x,y)$ in the sync code word and the channel output.
	The expected joint distribution, ${\mathsf P}$, induced by the sync pattern on the channel output, is defined as 
	\begin{equation*}
	\mathsf{P}_{\mathbf{s},\mathbf{y}}(x,y) \triangleq \mathsf{\hat{P}}_{\mathbf{s}}(x) Q(y|x)
	\end{equation*}
	where, $\mathsf{\hat{P}}_{\mathbf{s}}(x) = \frac{\mathsf{N}(x)}{N_A}, \text{ for all } x \in \mathcal{X}$ with $\mathsf{N}(x)$ denoting the number of occurrences of $x$ in the sync code word. 
	If the
	empirical distribution is close enough to the expected joint distribution , i.e., if $| \hat{\mathsf P} - {\mathsf P}| \leq \mu$ (for some $\mu > 0$), then, the decoder
	stops and declares $\hat{v} = t$.
	\begin{figure}
	\begin{center}
	\begin{tikzpicture}[scale=1,font=\scriptsize]
	\begin{axis}
	[	
	    axis y line=none,
	    axis x line=none,    
	    ymin=-1.2, ymax=3.2,
	    height = 4 cm, width = 5.5 cm,    
		title style={at={(axis cs:\xmid,-2.2)}},title=a. Error-less,font=\footnotesize,    
	]
		\def \xend{100}
		\def \xmid{50}
		\def \xbeg{1}
		\def \pbeg{71}
		\def \pend{80}
	    \addplot[ycomb,black!20,samples=\pbeg-1-\xbeg,domain=\xbeg:\pbeg-1] {rand};
	    \addplot[ycomb,black!70,samples=\pend-\pbeg,domain=\pbeg:\pend] {2+rand};
	    \addplot[ycomb,black!20,samples=\xend-\pend-1,domain=\pend+1:\xend] {rand};
	   	\pgfmathsetmacro{\ya}{\pbeg-.2}
	   	\pgfmathsetmacro{\yb}{\pend+.2}
	    \draw [black!70] (axis cs:\ya,-0.1) rectangle (axis cs:\yb,3);
	   	\pgfmathsetmacro{\yn}{\pbeg-1}
	    \node[below left , black,align=center,font=\scriptsize] at (axis cs:\yn,3) {Sync Packet};
	 \end{axis};
	 \end{tikzpicture}
	\begin{tikzpicture}[scale=1,font=\scriptsize]
	\begin{axis}
	[	
	    axis y line=none,
	    axis x line=none,    
	    ymin=-1.2, ymax=3.2,
	    height = 4 cm, width = 5.5 cm,    
		title style={at={(axis cs:\xmid,-2.2)}},title=b. Error E1,font=\footnotesize,    
	]
		\def \xend{100}
		\def \xmid{50}
		\def \xbeg{1}
		\def \pbeg{71}
		\def \pend{80}
		\def \nbeg{31}
		\def \nend{40}
	    \addplot[ycomb,black!20,samples=\nbeg-1-\xbeg,domain=\xbeg:\nbeg-1] {rand};
	    \addplot[ycomb,black!40,samples=\nend-\nbeg,domain=\nbeg:\nend] {1+rand};
	    \addplot[ycomb,black!20,samples=\pbeg-1-\nend-1,domain=\nend+1:\pbeg-1] {rand};
	    \addplot[ycomb,black!70,samples=\pend-\pbeg,domain=\pbeg:\pend] {2+rand};
	    \addplot[ycomb,black!20,samples=\xend-\pend-1,domain=\pend+1:\xend] {rand};
	   	\pgfmathsetmacro{\ya}{\pbeg-.2}
	   	\pgfmathsetmacro{\yb}{\pend+.2}
	    \draw [black!70] (axis cs:\ya,0.1) rectangle (axis cs:\yb,3);
	   	\pgfmathsetmacro{\na}{\nbeg-.2}
	   	\pgfmathsetmacro{\nb}{\nend+.2}
	    \draw [black!30] (axis cs:\na,0.1) rectangle (axis cs:\nb,3);
	   	\pgfmathsetmacro{\nc}{\nbeg-1}
	    \node[below left , black,align=center,font=\scriptsize]  at (axis cs:\nc,3)
		    {Sync-like\\noise};
	 \end{axis};
	 \end{tikzpicture}
	\begin{tikzpicture}[scale=1,font=\scriptsize]
	\begin{axis}
	[	
	    axis y line=none,
	    axis x line=none,    
	    ymin=-1.2, ymax=3.2,
	    height = 4 cm, width = 5.5 cm,    
	 title style={at={(axis cs:\xmid,-2.2)}},title=c. Error E2,font=\footnotesize,   
	]
		\def \xend{100}
		\def \xmid{50}
		\def \xbeg{1}
		\def \pbeg{71}
		\def \pend{80}
		\def \nbeg{65}
		\def \nend{74}
	    \addplot[ycomb,black!20,samples=\nbeg-1-\xbeg,domain=\xbeg:\nbeg-1] {rand};
	    \addplot[ycomb,black!40,fill=black!30,samples=\pbeg-1-\nbeg,domain=\nbeg:\pbeg-1] {1+rand};
	    \addplot[ycomb,black,fill=black!20,samples=\nend-\pbeg,domain=\pbeg:\nend] {2+rand};
	    \addplot[ycomb,black!70,fill=black!70,samples=\pend-\nend-1,domain=\nend+1:\pend] {2+rand};
	    \addplot[ycomb,black!20,samples=\xend-\pend-1,domain=\pend+1:\xend] {rand};
	   	\pgfmathsetmacro{\ya}{\pbeg-.2}
	   	\pgfmathsetmacro{\yb}{\pend+.2}
	    \draw [black!70] (axis cs:\ya,-0.2) rectangle (axis cs:\yb,3.1);
	   	\pgfmathsetmacro{\na}{\nbeg-.2}
	   	\pgfmathsetmacro{\nb}{\nend+.2}
	    \draw [black!30] (axis cs:\na,-0.1) rectangle (axis cs:\nb,3);
	   	\pgfmathsetmacro{\nc}{\nbeg-1}
	   	\node[below left , black,font=\scriptsize,align=center]  at (axis cs:\nc,3)
	    	{Sync-like\\Overlap};
	 \end{axis};
	 \end{tikzpicture}
	\begin{tikzpicture}[scale=1,font=\scriptsize]
	\begin{axis}
	[	
	    axis y line=none,
	    axis x line=none,    
	    ymin=-1.2, ymax=3.2,
	    height = 4 cm, width = 5.5 cm,    
	title style={at={(axis cs:\xmid,-2.2)}},title=d. Error E3,font=\footnotesize,    
	]
		\def \xend{100}
		\def \xmid{50}
		\def \xbeg{1}
		\def \pbeg{71}
		\def \pend{80}
	    \addplot[ycomb,black!20,samples=\pbeg-1-\xbeg,domain=\xbeg:\pbeg-1] {rand};
	    \addplot[ycomb,black!40,samples=\pend-\pbeg,domain=\pbeg:\pend] {1+3*rand};
	    \addplot[ycomb,black!20,samples=\xend-\pend-1,domain=\pend+1:\xend] {rand};
	   	\pgfmathsetmacro{\ya}{\pbeg-.2}
	   	\pgfmathsetmacro{\yb}{\pend+.2}
	    \draw [black!30] (axis cs:\ya,-1.1) rectangle (axis cs:\yb,3.1); 
	   	\pgfmathsetmacro{\ya}{\pbeg-1}
	    \node[below left , black,align=center,font=\scriptsize] at (axis cs:\ya,3.1)
	      {Noise-like \\sync};
	 \end{axis};
	 \end{tikzpicture}
	\end{center}
	\caption{ Examples for each partition of the error event $\{\hat{a} \neq a\}$ during packet reception}
	\label{fig:pkt_errors}
	\end{figure}
	
	The error event $\{ \hat{v} \neq v \}$ can be partitioned as discussed below (and as illustrated
	in the Figure~\ref{fig:pkt_errors}).
	\begin{itemize}
	    \item $E_1 : \hat{v} \in  \{1, \cdots, v-N\} \cup \{v+1, \cdots, A\}$. This corresponds to the event that the output symbols generated entirely by the zero input $x(0)$ is jointly typical.
	    \item $E_2 : \hat{v} \in \{v-N+1, \cdots, v-1\}$. This corresponds to the event that the output symbols generated partially by $x(0)$ and sync word is jointly typical.
		\item $E_3 : \hat{v} \notin \{ v \}$. This corresponds to the event that the output symbols generated by the sync word is not jointly typical. 
	\end{itemize}
	We note that the event $E_1 \cup E_2$ does not contain the event $E_3$ as we consider sequential
	frame detection.
	Using a union bound, we get,
	\[ {\mathsf P}(\{ \hat{v} \neq v \}) \leq {\mathsf P}(E_1) + {\mathsf P}(E_2) + {\mathsf P}(E_3) \]
	We will now show that $P(E_1), P(E_2)$ and $P(E_3)$ tend to zero for the listed conditions.
	Suppose that $A < e^{N_A \alpha(Q_A)}$ for all $A$. 
	Now, consider an asynchronism length of $\bar{A}=e^{N_A \epsilon_1 (\alpha(Q_A) - \epsilon_2)}$, where
	$0 < \epsilon_1 < 1$ and $\epsilon_2 > 0$. Clearly, $\bar{A}  < e^{N_A \alpha(Q_A)}$. 
	
	The probability that the channel output for an
	input sequence composed entirely of $x(0)$ is jointly typical can be computed as discussed in \cite{Chandar2008},
	\[ e^{-N_A(1-\frac{1}{K}) [ D( Q_A(\cdot|x(1)) || Q_A(\cdot|x(0)) ) + H(Q_A(\cdot|x(1))) - \delta]} \]
	where $\delta$ is a function of $\mu$ (of the joint typicality decoder) and tends to zero as $\mu \rightarrow 0$.
	Substituting for $D( Q_A(\cdot|x(1)) || Q_A (\cdot|x(0)) ) = \alpha(Q_A)$ in the above expression and ignoring the non-negative entropy term, we get
	the following upper bound
	\[ e^{-N_A(1-\frac{1}{K}) [ \alpha(Q_A) - \delta]} \]
	We can now bound ${\mathsf P}(E_1)$ using a union bound as,
	\begin{eqnarray*}
	{\mathsf P}(E_1) &\leq& \bar{A} \times e^{-N_A (1-\frac{1}{K}) (\alpha(Q_A) - \delta)}\\
	 &=& e^{N_A \epsilon_1(\alpha(Q_A) - \epsilon_2) }e^{-N_A(1-\frac{1}{K}) (\alpha(Q_A) - \delta)}\\
	&=& e^{N_A \alpha(Q_A) (\epsilon_1 - (1 - \frac{1}{K}))} e^{-N_A (\epsilon_1 \epsilon_2 - \delta)}
	\end{eqnarray*}
	For large $K$, small $\delta$ (depends on $\mu$ of the typicality decoder) and as $A \rightarrow \infty$ (and hence,
	$N_A \alpha(Q_A) \rightarrow \infty$), we have, ${\mathsf P}(E_1) \rightarrow 0$ for any $\epsilon_1$ and $\epsilon_2$.
	Thus $\mathsf{P}(E_1) \rightarrow 0$ for all $A < e^{N_A \alpha(Q_A)}$
	
	The probability that the channel output for an
	input sequence composed partially of $x(0)$ and $x(1)$ (a shifted sequence) is jointly typical can be upper bounded as
	\begin{eqnarray*}
	&e^{-N_A \frac{1}{K} [ D( Q(\cdot|x(1)) || Q (\cdot|x(0)) ) + H(Q(\cdot|x(1))) - \delta]}&\\
	&\leq  e^{-N_A \frac{1}{K} [ \alpha(Q_A) - \delta]}&
	\end{eqnarray*} 
	Using a union bound for ${\mathsf P}(E_2)$, we get,
	\[ {\mathsf P}(E_2) \leq N_A e^{-N_A \frac{1}{K} [ \alpha(Q_A) - \delta]} \]
	For small $\delta$ and as $A \rightarrow \infty$ (and hence, $N_A \alpha(Q_A) \rightarrow \infty$), we have ${\mathsf P}(E_2) \rightarrow 0$.
	
	If $N_A \rightarrow \infty$ as $A \rightarrow \infty$, then ${\mathsf P}(E_3) \rightarrow 0$ (follows from the
	weak law of large numbers).
	
	Thus, we have ${\mathsf P}(\{\hat{v} \neq v\}) \rightarrow 0$ if $e^{N_A \alpha(Q_A)} > A$.
	The above arguments show the achievability of asymptotic error-free frame synchronisation
	if $e^{\alpha(Q_A) N_A} > A$.
	The converse follows directly from the discussion in \cite{Chandar2008}, if we consider the special case of $\alpha(Q_A) = \alpha(Q)$ and 
	let $N_A \rightarrow \infty$.
	\end{proof}
\subsection{Proof of Corollary~\ref{cor:energy}}
	\begin{proof}
	From \cite{Chandar2008}, we know that $\alpha(Q_A) = \frac{P_A}{2 \sigma^2}$ for an AWGN channel.
	From Theorem~\ref{thm:generalization}, we have ${\mathsf {P_A}}(\{\hat{v} \neq v\}) \rightarrow 0$
	if $e^{\alpha(Q_A) N_A} = e^{\frac{P_A}{2\sigma^2}N_A}> A$.
	\end{proof}
\fi 
\end{document}